\documentclass[aip,apl,amsmath,amssymb,reprint]{revtex4-1}

\usepackage{graphicx}
\usepackage{dcolumn}
\usepackage{bm}

\usepackage[utf8]{inputenc}
\usepackage[T1]{fontenc}
\usepackage{mathptmx}
\usepackage{etoolbox}
\usepackage{gensymb}
\usepackage{xfrac}
\usepackage{xcolor}

\makeatletter
\def\@email#1#2{%
 \endgroup
 \patchcmd{\titleblock@produce}
  {\frontmatter@RRAPformat}
  {\frontmatter@RRAPformat{\produce@RRAP{*#1\href{mailto:#2}{#2}}}\frontmatter@RRAPformat}
  {}{}
}%
\makeatother
\begin{document}

\title{Direct Fabrication of a Superconducting Two-Dimensional Electron Gas on KTaO$_3$(111) via Mg-Induced Surface Reduction}
\author{Chun Sum Brian Pang}
\author{Bruce A. Davidson}
\author{Fengmiao Li}
\author{Mohamed Oudah}
\author{Peter C. Moen}
\author{Steef Smit}
\affiliation{Quantum Matter Institute, University of British Columbia, Vancouver, British Columbia V6T 1Z4, Canada}
\affiliation{Department of Physics \& Astronomy, University of British Columbia, Vancouver, British Columbia V6T 1Z1, Canada}

\author{Cissy T. Suen}
\affiliation{Quantum Matter Institute, University of British Columbia, Vancouver, British Columbia V6T 1Z4, Canada}
\affiliation{Department of Physics \& Astronomy, University of British Columbia, Vancouver, British Columbia V6T 1Z1, Canada}
\affiliation{Max Planck Institute for Solid State Research, 70569 Stuttgart, Germany}

\author{Simon Godin}
\affiliation{Quantum Matter Institute, University of British Columbia, Vancouver, British Columbia V6T 1Z4, Canada}
\affiliation{Department of Physics \& Astronomy, University of British Columbia, Vancouver, British Columbia V6T 1Z1, Canada}

\author{Sergey A. Gorovikov}
\author{Marta Zonno}
\affiliation{Canadian Light Source, Inc., Saskatoon, Saskatchewan S7N 2V3, Canada}

\author{Pinder Dosanjh}
\author{Sergey Zhdanovich}
\author{Giorgio Levy}
\author{Matteo Michiardi}
\author{Alannah M. Hallas}
\author{George A. Sawatzky}
\affiliation{Quantum Matter Institute, University of British Columbia, Vancouver, British Columbia V6T 1Z4, Canada}
\affiliation{Department of Physics \& Astronomy, University of British Columbia, Vancouver, British Columbia V6T 1Z1, Canada}

\author{Robert J. Green}
\affiliation{Quantum Matter Institute, University of British Columbia, Vancouver, British Columbia V6T 1Z4, Canada}
\affiliation{Department of Physics \& Astronomy, University of British Columbia, Vancouver, British Columbia V6T 1Z1, Canada}
\affiliation{Department of Physics \& Engineering Physics, University of Saskatchewan, Saskatoon, Saskatchewan S7N 5E2, Canada}

\author{Andrea Damascelli}
\author{Ke Zou}
\affiliation{Quantum Matter Institute, University of British Columbia, Vancouver, British Columbia V6T 1Z4, Canada}
\affiliation{Department of Physics \& Astronomy, University of British Columbia, Vancouver, British Columbia V6T 1Z1, Canada}

\email{csbpang@phas.ubc.ca, damascelli@physics.ubc.ca, kzou@phas.ubc.ca}
\date{\today}

\begin{abstract}
Two-dimensional electron gases (2DEGs) at the surfaces of KTaO$_3$ have become an exciting platform for exploring strong spin-orbit coupling, Rashba physics, and low-carrier-density superconductivity. Yet, a large fraction of reported KTaO$_3$-based 2DEGs has been realized through chemically complex overlayers that both generate carriers and can obscure the native electronic structure, making spectroscopic access to the underlying 2DEG challenging. Here, we demonstrate a simple and direct method to generate a superconducting 2DEG on KTaO$_3$(111) using Mg-induced surface reduction in molecular-beam epitaxy (MBE). Mg has an extremely low sticking coefficient at elevated temperatures, enabling the formation of an ultrathin (<1--2 monolayer) MgO layer that is transparent to soft X-ray photoemission (XPS) and angle-resolved photoemission spectroscopy (ARPES). This allows direct measurement of the surface chemistry and low-energy electronic structure of the pristine reduced surface without the need for a several-nanometer-thick capping layer. XPS shows clear reduction of Ta$^{5+}$ to lower oxidation states, while ARPES reveals a parabolic Ta 5d conduction band with a $\sim$150~meV bandwidth and additional subband features arising from quantum confinement. Transport measurements confirm a superconducting transition below 0.7~K. Together, these results demonstrate a chemically straightforward and controllable pathway for fabricating spectroscopically accessible superconducting 2DEGs on KTaO$_3$(111), and provide a powerful new platform for investigating the mechanisms underlying orientation-dependent superconductivity in KTaO$_3$-based oxide interfaces. 
\end{abstract}

\maketitle

KTaO$_3$ (KTO) is a wide-gap perovskite insulator ($E_g \approx 3.6$ eV\cite{jellison_optical_2006}) whose surfaces can host two-dimensional electron gases (2DEGs) with remarkable physical properties. Conductive states have been engineered through \textit{in situ} cleaving\cite{king_subband_2012, santander-syro_orbital_2012, bareille_two-dimensional_2014} or by growing overlayers such as AlO$_x$, EuO, LaTiO$_3$, and LaVO$_3$.\cite{zou_latio3ktao3_2015, mallik_superfluid_2022, chen_orientation-dependent_2024, liu_superconductivity_2023} These methods induce the formation of oxygen vacancies and/or atomic substitution at potassium sites near the KTO surface.\cite{gupta_ktao3new_2022, liu_two-dimensional_2021, mallik_superfluid_2022} Consequently, Ta$^{5+}$ ions are reduced to lower oxidation states such as Ta$^{4+}$, leading to increased electron occupation in the Ta 5d conduction band and the formation of 2DEGs.\cite{gupta_ktao3new_2022, mallik_superfluid_2022}

Due to the presence of heavy Ta atoms, KTO exhibits strong spin–orbit coupling of about 0.3 eV, which gives rise to pronounced Rashba spin splitting and enables efficient spin–charge interconversion in KTO-based 2DEGs.\cite{mattheiss_energy_1972,vicente-arche_spincharge_2021,varotto_direct_2022,al-tawhid_spin--charge_2025} Beyond their spintronic functionality, several KTO 2DEGs formed by overlayer growth have also been found to exhibit two-dimensional superconductivity, with the critical temperature ($T_c$) strongly dependent on crystallographic orientation, reaching up to 2.2 K on KTO(111), 1.3 K on KTO(110), and only 0.25 K on KTO(001).\cite{gupta_ktao3new_2022,liu_two-dimensional_2021,mallik_superfluid_2022,chen_orientation-dependent_2024,liu_superconductivity_2023,hua_tunable_2022,chen2025czo_kto001} The underlying mechanism of the orientation-dependent superconductivity observed in KTO-based 2DEGs remains a topic of active discussion.\cite{liu_tunable_2023, chen_orientation-dependent_2024} 

To develop a microscopic understanding of this behavior, it is essential to directly resolve the intrinsic electronic structure of KTO-based 2DEGs under well-controlled conditions. However, many established methods for creating such 2DEGs rely on chemically complex or relatively thick overlayers to drive the necessary redox processes. These overlayers can obscure the 2DEG from surface-sensitive probes and may introduce structural or chemical perturbations, making it challenging to compare different orientations under identical surface conditions. This motivates the development of alternative approaches that produce a clean, minimally perturbed, and spectroscopically accessible superconducting 2DEG directly at the KTO surface.

Here, we demonstrate that a superconducting 2DEG can be realized on the KTO(111) surface through direct Mg reduction, enabling \textit{in situ} access to both the surface chemistry and low-energy electronic structure via X-ray photoemission spectroscopy (XPS) and angle-resolved photoemission spectroscopy (ARPES). Using molecular beam epitaxy (MBE), we developed a three-stage growth procedure to fabricate an MgO/KTO(111) heterostructure that introduces a 2DEG at the KTO(111) surface. Reflection high-energy electron diffraction (RHEED) was employed to monitor the surface structure and guide each stage of the fabrication process.

\begin{figure*}
\centering
\includegraphics[width=\textwidth]{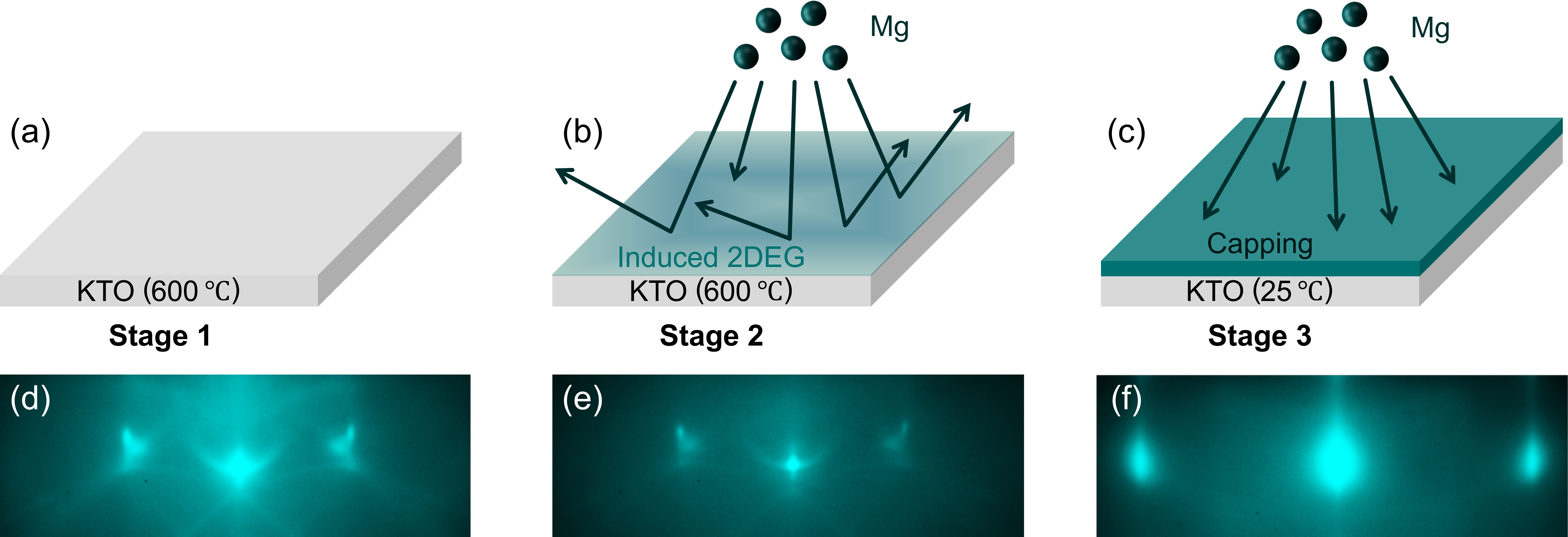}
\caption{\label{KTO_steps}Fabrication of the MgO/KTO(111) sample. Schematic diagrams of (a) Stage~1, substrate degassing; (b) Stage~2, high-temperature surface reduction using Mg; and (c) Stage~3, room-temperature capping. The corresponding RHEED patterns along the $[\bar{1}\bar{1}2]$ direction are shown in (d) to (f).}
\end{figure*}

Stage~1 involves substrate degassing. KTO(111) single-crystal substrates (10~$\times$~10~$\times$~0.5~mm$^3$, MTI Corporation) were cleaved into 2.5~$\times$~2.5~$\times$~0.5~mm$^3$ pieces using a diamond scribing pen. Each substrate was degassed at 600~\textdegree C for 1~hour under ultra-high vacuum (UHV) conditions better than $2.0 \times 10^{-8}$~Torr [Fig.~\ref{KTO_steps}(a)]. The RHEED pattern, which repeats every $60^\circ$ of sample rotation (not shown) as expected for the KTO(111) surface, remains unchanged during heating and degassing aside from slight blurring [Fig.~\ref{KTO_steps}(d)], suggesting the absence of structural reconstruction of the KTO(111) surface at elevated temperatures. The mild reduction in sharpness may be attributed to the formation of defects at high temperature, a behavior previously observed by atomic force microscopy on KTO(001) surfaces cleaved and annealed in UHV.\cite{alexander_atomic-scale_2024}

Stage~2 involves the high-temperature surface reduction of the KTO substrate using Mg. The substrate was held at 600~\textdegree C and exposed to an elemental Mg flux of $2.5 \times 10^{13}\;\mathrm{atoms\;cm^{-2}\,s^{-1}}$ for 12~minutes, with the Mg effusion cell maintained at 310~\textdegree C [Fig.~\ref{KTO_steps}(b)]. Since the substrate temperature is much higher than that of the Mg cell, the incident Mg atoms have a very low sticking coefficient and are largely re-evaporated or scattered upon arrival. Nevertheless, a fraction of the incoming Mg atoms reacts with lattice oxygen to form MgO, which has a much higher sticking coefficient and adsorbs onto the KTO surface as an ultrathin layer or as small clusters. This reaction extracts oxygen from the topmost KTO unit cells, creating oxygen vacancies that drive electron transfer into the Ta 5d conduction bands and reduce surface Ta$^{5+}$ ions. As a result, a 2DEG is stabilized at the KTO(111) surface after reduction, and the corresponding RHEED pattern dims but remains visible, with no additional diffraction spots or streaks observed [Fig.~\ref{KTO_steps}(e)]. Since the RHEED electron beam probes only the top few atomic layers, the similarity of diffraction patterns before and after reduction indicates that the KTO(111) surface is not significantly obscured by the adsorbed MgO. This suggests that the MgO overlayer is no more than 1--2 monolayers thick (i.e., $< 1$~nm), consistent with the low sticking coefficient of Mg on a hot substrate. Only Mg atoms that react with the KTO surface to form MgO could contribute to film formation; with the resulting ultrathin overlayer, the 2DEG remains highly exposed, allowing direct \textit{in situ} probing of its electronic structure using surface-sensitive techniques such as XPS and ARPES immediately following Stage~2.

Stage~3 involves the growth of a capping layer at room temperature, which can be performed either directly after surface reduction (Stage~2) or following \textit{in situ} XPS and ARPES measurements. The sample was allowed to settle to room temperature, then exposed again to an elemental Mg flux of $2.5 \times 10^{13}\;\mathrm{atoms\;cm^{-2}\,s^{-1}}$ for 12~minutes, with the Mg cell maintained at 310~\textdegree C [Fig.~\ref{KTO_steps}(c)]. At this lower substrate temperature, Mg atoms exhibit a high sticking coefficient and readily deposit on top of the 2DEG, forming an Mg overlayer. Based on the flux and deposition time, the Mg overlayer thickness is estimated to be approximately 4~nm. The RHEED pattern [Fig. 1(f)] transforms into a new set of distinct spots with increased spacing, indicative of a crystalline hcp Mg capping layer. The broadening of the diffraction spots suggests finite lateral coherence and a rough, terraced surface morphology.\cite{ichimiya_reflection_2004} The ratio of RHEED spot spacing between Fig.~\ref{KTO_steps}(d) and Fig.~\ref{KTO_steps}(f) is 0.57:1, matching the inverse ratio of the in-plane lattice constants of KTO(111) (5.62~\AA) and Mg (3.21~\AA). The pattern also repeats every $60^\circ$, consistent with the hexagonal close-packed structure of Mg [Fig.~S1]. Upon air exposure, the Mg overlayer oxidizes into MgO, serving as a protective capping layer for \textit{ex situ} transport measurements of the 2DEG.

As a control experiment, the same three-stage fabrication procedure was applied to a sapphire substrate; the sample remains insulating, confirming the absence of a 2DEG and indicating that the Mg capping layer fully oxidizes into MgO. A separate control sample of KTO subjected only to Stage 1 (substrate degassing) is also insulating, confirming that UHV annealing at 600~\textdegree C does not introduce detectable oxygen vacancies and primarily serves to degas the substrate. In addition, a KTO sample fabricated without Stage~2 (i.e., without high-temperature Mg exposure) is likewise insulating. These results demonstrate that the electronic properties of the MgO/KTO(111) heterostructure produced by the three-stage approach arise directly from the 2DEG formed at the reduced KTO(111) surface.

\begin{figure}[t]
\centering
\includegraphics[width=0.9\columnwidth]{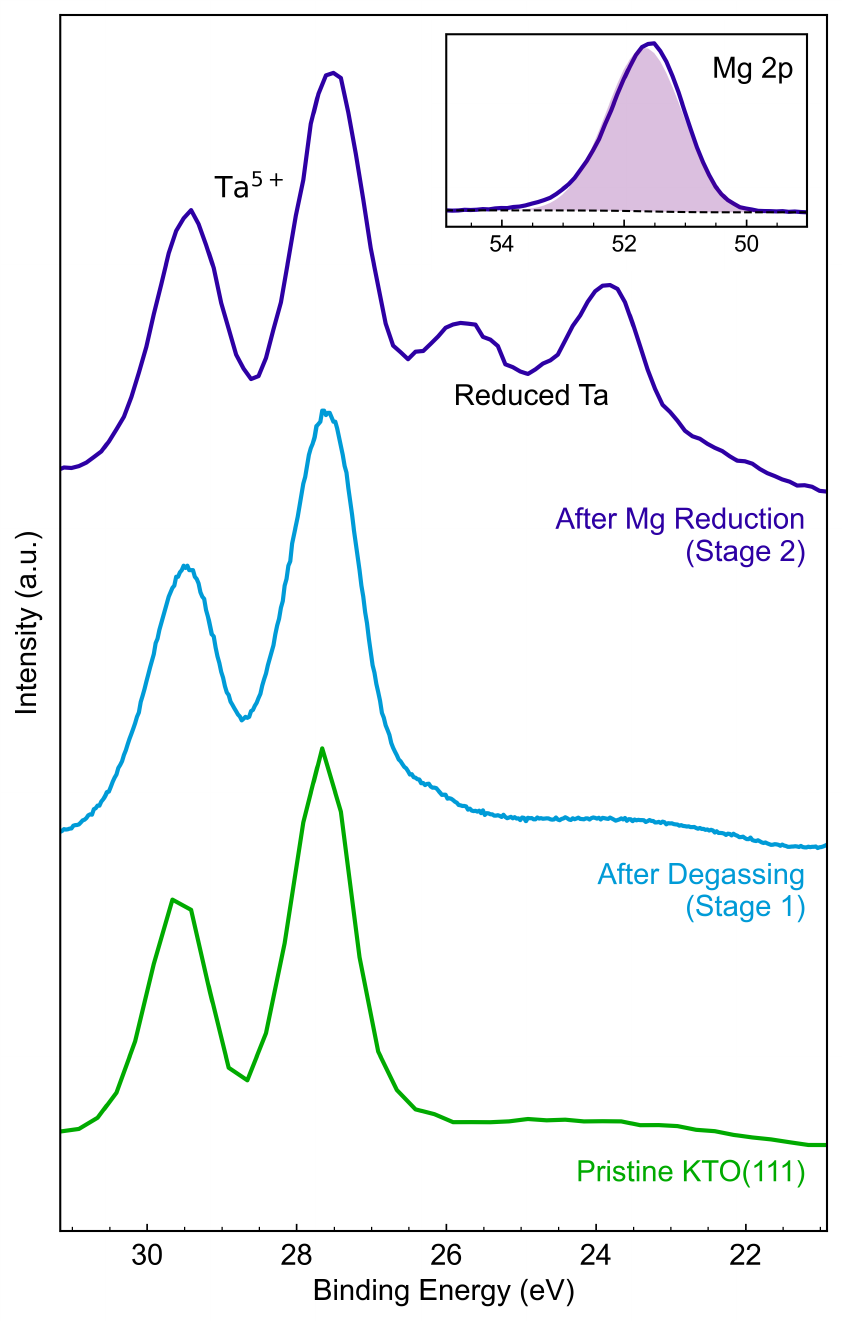}
\caption{\label{XPS}XPS spectra of an MgO/KTO(111) sample. The Ta 4f core-level spectra were acquired from a pristine KTO(111) substrate surface (green), after degassing (light blue), and after Mg-induced surface reduction (dark blue). The inset shows the Mg 2p core-level spectrum after the reduction. The symmetric Gaussian fit (light purple) is centered at 51.67 eV with a FWHM of 1.46 eV, and the Shirley background is shown as a black dashed line.}
\end{figure}

To examine the surface chemistry associated with the Mg treatment, low-energy X-ray photoemission spectroscopy (XPS) measurements were performed around the Ta 4f and Mg 2p core levels before and after Mg exposure [Fig.~\ref{XPS}]. The measurements were carried out at the Quantum Materials Spectroscopy Centre (QMSC) beamline of the Canadian Light Source (CLS), using a Scienta R4000 hemispherical analyzer with angular and energy resolutions better than 0.1\degree and 15~meV, respectively. The samples were measured at a temperature of 15~K and under a base pressure below $5.0 \times 10^{-11}$~Torr. The spectra were acquired at normal emission with p-polarized light at a photon energy of 120~eV. The emitted photoelectrons have kinetic energies of 60--100~eV, yielding inelastic mean free paths of $\sim 5~\text{\AA}$ in oxides.\cite{seah_quantitative_1979} At normal emission, this gives a 95\% information depth of $\sim$1.5~nm ($\approx$3$\lambda$),\cite{Powell2020IMFP} indicating that the measurements probe only the top $\sim$3--4~unit cells and are therefore highly surface sensitive. The binding-energy scale was calibrated by fitting the Fermi edge of a freshly prepared gold surface mounted on the cryostat and electrically connected to the sample holder.

The pristine KTO(111) surface is dominated by Ta$^{5+}$ species, as evidenced by the Ta 4f doublet centered at 27.6 and 29.5~eV [Fig.~\ref{XPS}, green curve].\cite{moulder_handbook_1992,brumbach_evaluating_2014,Simpson2017XPS} After substrate degassing (Stage~1), there is no observable change in the Ta 4f spectrum [Fig.~\ref{XPS}, light blue curve], indicating that the surface remains largely unreduced. In contrast, following the Mg surface treatment (Stage~2), the Ta 4f spectrum displays a pronounced increase in low-binding-energy spectral weight relative to the Ta$^{5+}$ doublet [Fig.~\ref{XPS}, dark blue curve]. Because 5d Ta oxides exhibit strong covalency and core-level final-state screening can produce energy shifts comparable to nominal chemical shifts, it is highly nontrivial to uniquely decompose the reduced spectral weight into discrete formal oxidation states (Ta$^{4+}$/Ta$^{3+}$/Ta$^{2+}$). We therefore refer to this additional intensity as reduced Ta (Ta$^{<5+}$).\cite{brumbach_evaluating_2014,Simpson2017XPS} This assignment-independent increase in reduced Ta spectral weight is consistent with electron accumulation at the KTO surface, signaling the formation of a surface 2DEG. This direct spectroscopic signature is a key advantage of our ultrathin-overlayer approach.

\begin{figure}[b]
\centering
\includegraphics[width=\columnwidth]{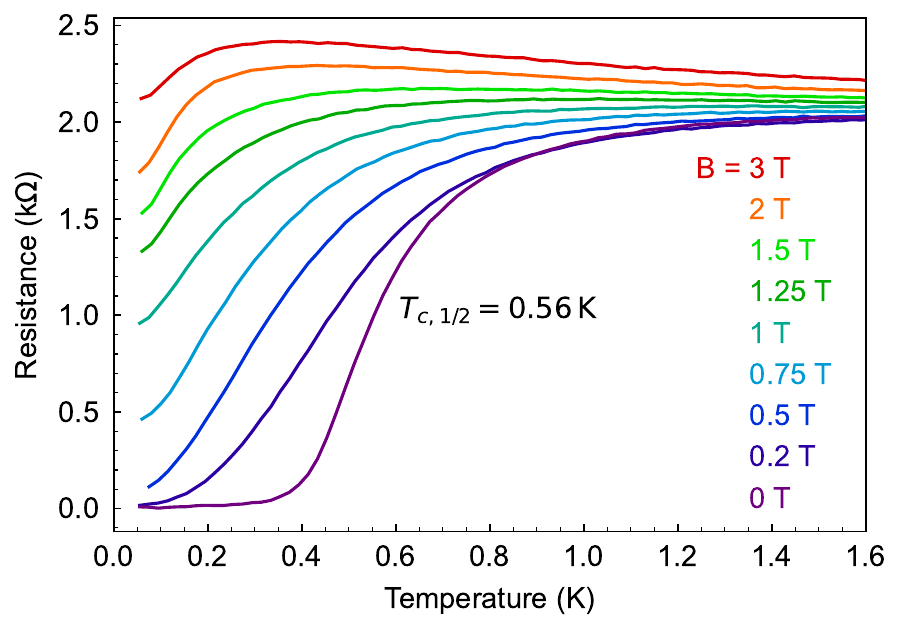}
\caption{\label{SC}Low-temperature transport measurements. Resistance as a function of temperature between 60~mK and 1.6~K for out-of-plane magnetic fields ranging from 0 to 3~T. The midpoint critical temperature ($T_{\mathrm{c,1/2}}$), defined at half the normal-state resistance, is shown.}
\end{figure}

In addition, after Mg-induced surface reduction (Stage~2), a strong Mg 2p peak is observed [Fig.~\ref{XPS}, inset]. The peak can be approximated by a symmetric Gaussian centered at 51.67~eV with a full width at half maximum (FWHM) of 1.46~eV. The binding energy is significantly higher than that of metallic Mg ($\sim$49.5 eV) and lies within the reported range for MgO ($\sim$50.8--51.8 eV).\cite{Skaanvik2025MgXPS,CristMgXPSDatabase} Furthermore, the near-symmetric Gaussian profile contrasts with the asymmetric Doniach--\v{S}unji\'c line shape expected for metallic Mg, and the observed FWHM is also consistent with reported MgO spectra.\cite{Skaanvik2025MgXPS,CristMgXPSDatabase} These features indicate that the deposited Mg is predominantly oxidized rather than metallic. This observation is consistent with oxidation of Mg via an interfacial redox process with lattice oxygen at the KTO surface, driving the formation of oxygen vacancies, the reduction of Ta, and consequently the formation of a surface 2DEG.

To investigate the possible realization of superconductivity induced by Mg reduction, we performed \textit{ex situ} temperature-dependent transport measurements on the MgO/KTO(111) samples [Fig.~\ref{SC}]. The samples used for transport measurements were fabricated following the procedure described above but were not exposed to synchrotron irradiation prior to transport measurements. The MgO capping layers were lightly scratched near the four corners to expose the 2DEGs. Then, the 2DEGs were immediately contacted with gold wires using indium cold pressing in a van der Pauw geometry. The transport measurements were performed using the dilution refrigerator option of a Quantum Design Physical Property Measurement System.

A superconducting phase transition is observed, with an onset temperature ($T_{\mathrm{c,onset}}$) of 0.73~K and the resistance reaching zero ($T_{\mathrm{c,0}}$) at 0.23~K [Fig.~S2]. The systematic suppression of the transition temperature under applied out-of-plane magnetic fields further confirms the superconducting nature of this transition. At zero field, the midpoint critical temperature is $T_{\mathrm{c,1/2}} = 0.56$~K, lower than the 2.2~K reported by Liu \textit{et al.}\cite{liu_two-dimensional_2021} The origin of this difference remains unclear and may reflect differences in carrier density, disorder, or details of the reduction process. Since Mg, MgO, and bulk KTO are all non-superconducting, we conclude that the observed superconductivity originates from the 2DEG formed at the Mg-reduced KTO(111) surface.

\begin{figure}
\centering
\includegraphics[width=\columnwidth]{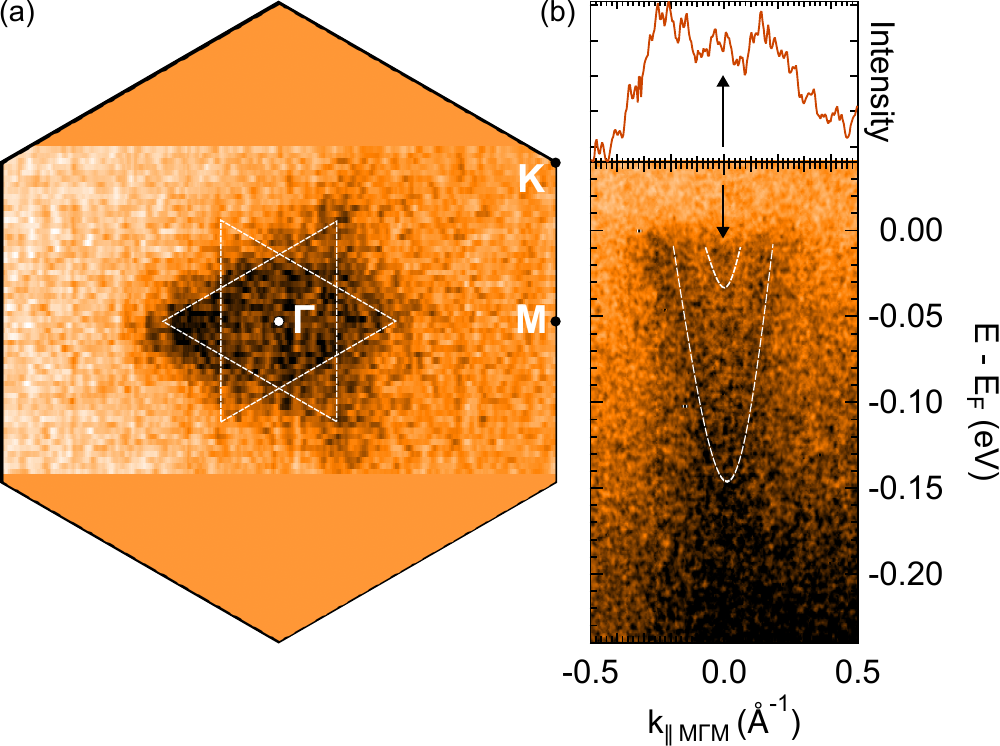}
\caption{\label{ARPES_data}ARPES data of the Mg-reduced KTO(111) surface. (a) Fermi surface in the first surface Brillouin zone with high-symmetry points labeled. (b) Band dispersion along the M--$\Gamma$--M direction, with the momentum distribution curve (MDC, top) at $E_F$. White dotted lines are included in both the Fermi surface and band-dispersion panels as guides to the eye.}
\end{figure} 

To probe the electronic structure of the 2DEG in the MgO/KTO(111) system, we performed \textit{in situ} ARPES measurements on the Mg-reduced KTO(111) surface immediately after Stage~2. The measurements were carried out at the QMSC beamline at CLS on the same endstation used for the XPS. The samples were aligned with the $\Gamma$--M direction parallel to the analyzer slit, then measured at a temperature of 15~K and under a base pressure below $5.0 \times 10^{-11}$~Torr. The ARPES spectra were acquired with p-polarized light at a photon energy of 105.5~eV.

The resulting Fermi surface [Fig.~\ref{ARPES_data}(a)] exhibits a predominantly triangular shape with small protrusions located at the midpoints of the triangle edges. The three corners of the triangle and the three protrusions alternate upon rotation around $\Gamma$, and they point toward the M points of the surface Brillouin zone. This pattern is consistent with the star-of-David–like Fermi surfaces reported by Bruno \textit{et al.}\cite{bruno_band_2019} on free KTO(111) surfaces and by Mallik \textit{et al.}\cite{mallik_electronic_2023} on Eu/KTO(111) 2DEGs. As discussed in Ref.~\citenum{bruno_band_2019}, such a star-shaped Fermi surface originates from hybridization among the three Ta 5d $t_{2g}$ orbitals, further supporting that the 2DEG in MgO/KTO(111) is generated through the reduction of surface Ta$^{5+}$ ions by Mg.

The corresponding band dispersion along the M--$\Gamma$--M direction exhibits a parabolic shape with an occupied bandwidth of approximately 150~meV [Fig.~\ref{ARPES_data}(b)]. In addition, a shallow subband-like feature appears near the $\Gamma$ point and manifests as a central peak in the momentum distribution curve (MDC) at the Fermi level [Fig.~\ref{ARPES_data}(b), top]. This feature has been described as a quantum well state---a replica of the main parabolic band arising from quantum confinement of the 2DEG at the KTO surface.\cite{bruno_band_2019,mallik_electronic_2023} The presence of this subband in our dispersion data is therefore consistent with the expected two-dimensional nature of the electronic states at the reduced KTO(111) surface.

In summary, we have demonstrated a simple, direct, and generalizable MBE-based method for generating superconducting 2DEGs on KTO(111) via Mg-induced surface reduction while retaining full \textit{in situ} spectroscopic access. \textit{Ex situ} magnetic-field-dependent transport measurements confirm the superconducting transition to a zero-resistance ground state, and its connection to a confined 2DEG at the Mg-reduced KTO(111) surface is established through the observation of quantum-well subband features in \textit{in situ} ARPES.

This approach can be extended to KTO(110) and KTO(001), enabling direct investigation of the surface chemistry and electronic structure of 2DEGs across different crystallographic orientations. Such studies may provide valuable insight into the origin of the unconventional, orientation-dependent superconductivity observed in KTO-based heterostructures. Moreover, the versatility and simplicity of the Mg-reduction method make it a promising strategy for generating 2DEGs in other insulating oxides, broadening the available toolkit for oxide interface engineering.

See the Supplementary Material for additional details on RHEED, ARPES, and transport measurements.

This research was undertaken thanks in part to funding from the Max Planck–UBC–UTokyo Centre for Quantum Materials and the Canada First Research Excellence Fund, Quantum Materials and Future Technologies. This project is also funded by the Natural Sciences and Engineering Research Council of Canada (NSERC); the Canada Foundation for Innovation (CFI); the Department of National Defence (DND); the British Columbia Knowledge Development Fund (BCKDF); the Mitacs Accelerate Program; the QuantEmX Program of the Institute for Complex Adaptive Matter; the Gordon and Betty Moore Foundation’s EPiQS Initiative, Grant GBMF4779 to A.D..; the Canada Research Chairs Program (A.D.); and the CIFAR Quantum Materials Program (A.D.). Use of the Canadian Light Source (Quantum Materials Spectroscopy Centre), a national research facility of the University of Saskatchewan, is supported by CFI, NSERC, the National Research Council, the Canadian Institutes of Health Research, the Government of Saskatchewan and the University of Saskatchewan.

\section*{Author Declarations}
\subsection*{Conflict of Interest}
The authors have no conflicts to disclose.

\subsection*{Author Contributions}
\textbf{Chun Sum Brian Pang}: Conceptualization (equal); Data curation (lead); Formal analysis (lead); Investigation (lead); Methodology (lead); Software (equal); Writing – original draft (lead); Writing – review and editing (lead). 
\textbf{Bruce A. Davidson}: Conceptualization (equal); Investigation (supporting); Methodology (supporting); Validation (equal); Writing – review and editing (equal).
\textbf{Fengmiao Li}: Conceptualization (equal); Methodology (supporting).
\textbf{Mohamed Oudah}: Investigation (supporting). 
\textbf{Peter Moen}: Investigation (supporting). 
\textbf{Steef Smit}: Investigation (supporting). 
\textbf{Cissy T. Suen}: Investigation (supporting). 
\textbf{Simon Godin}: Formal analysis (supporting); Software (equal). 
\textbf{Sergey A. Gorovikov}: Investigation (supporting). 
\textbf{Marta Zonno}: Investigation (supporting). 
\textbf{Pinder Dosanjh}: Investigation (supporting); Methodology (supporting).
\textbf{Sergey Zhdanovich}: Validation (equal).
\textbf{Giorgio Levy}: Validation (equal).
\textbf{Matteo Michiardi}: Validation (equal).
\textbf{Alannah M. Hallas}: Supervision (supporting).
\textbf{George A. Sawatzky}: Supervision (supporting); Validation (equal).
\textbf{Robert J. Green}: Formal analysis (supporting); Software (equal); Validation (equal); Writing – review and editing (equal).
\textbf{Andrea Damascelli}: Conceptualization (equal); Funding acquisition (equal); Project administration (equal); Resources (equal); Supervision (equal); Validation (equal); Writing – review and editing (equal).
\textbf{Ke Zou}: Conceptualization (lead); Funding acquisition (equal); Project administration (equal); Resources (equal); Supervision (lead); Validation (equal); Writing – review and editing (equal).

\section*{Data Availability}
The data that support the findings of this study are available from the corresponding author upon reasonable request.

\section*{References}
\bibliographystyle{aipnum4-1}
\bibliography{ref}

\end{document}